# Luminescent Waveguides In-situ Integrated with Organic Solar Cells for Internet of Things

Sadra Sadeghi[1,*], Mertcan Han[2,*], Shashi Bhushan Srivastava[2,*] and Sedat Nizamoglu[1, 2, ϕ]

[1] Graduate School of Materials Science and Engineering, Koc University, Istanbul, Turkey, 34450.

[2] Department of Electrical and Electronics Engineering, Koc University, Istanbul, Turkey, 34450.

ϕ Corresponding author: snizamoglu@ku.edu.tr

* These authors contributed equally.

**ABSTRACT**

Transparent electronics offer exciting light harvesting solutions for generation of electrical power via demonstration of 'see-through' optoelectronic devices. The wireless energy harvesting ability can empower unplugged and battery-free operations for Internet of Things (IoT) devices. In this study, we report a transparent, luminescent, and elastomeric optical waveguide incorporating quantum dots that is in-situ coupled with organic solar cell array made of $P3HT:PC_{61}BM$ bulk heterojunction. CdSe@ZnS QDs have a photoluminescence quantum yield (PLQY) of 91% and are synthetically engineered to match their photoluminescence spectra with the photo-response of $P3HT:PC_{61}BM$ solar cells for efficient energy harvesting. Integrated devices can generate sufficient power for the active radio frequency identification (RFID) tags to send signals in the communication distance of 35 meters at low illumination level of 0.1-sun and ~0.1 km under 1-sun condition, respectively, which is sufficient for indoor and outdoor communications. Advantageously, the combination of organic solar cells with the waveguide during the elastomer curing leads to the elimination of the undesired post-fabrication processes such as alignment of the solar cells with waveguide and gluing the separate parts with curable polymers. This study paves the way toward using luminescence as transparent, efficient, and configurable energy harvesting solutions for IoT applications.

**INTRODUCTION**

Transparent electronics with both high transparency and excellent performance are emerging technologies for the next generation of 'see-through' functional devices [1]. In the transparent electronics technology, the efforts are focused to simultaneously produce 'invisible' electronic circuity and maintain the sufficient optical transparency in optoelectronic devices [2]. The most recently developed technologies of transparent electronics include organic light-emitting diodes (OLEDs) [3], thin film transistors (TFTs) [4], solar cells [5], supercapacitors [6], memory devices [7], and nanogenerators [8].

Luminescent solar concentrators (LSCs) are one of the most promising sunlight energy harvesting technology that can 'invisibly' be integrated inside the buildings in order to generate the sufficient electrical power for indoor use [9]. As a result, it can be categorized into the recent and innovative class of transparent electronics. The most common architectures of LSCs consist of a large-area transparent slab made of polymer or glass. The transparent slab incorporates fluorescent particles, which absorb the incoming light and re-emit luminescence at longer wavelengths. The emitted light is then waveguided through the slab and absorbed by the PV coupled at the edges. Such transparent slabs can be aesthetically integrated with a wide variety of objects for smart devices (i.e., internet of things (IoT)) with the ability to transfer data over a network. However, in general, light energy harvesting via LSCs are mostly focused on the coupling of luminescence to the conventional silicon solar cells [10-12] that operate efficiently at high-intensity radiation [13, 14]. Alternatively, organic photovoltaics are promising candidates for the next-generation renewable energy harnessing technologies due to solution-processability, large-scale production and flexibility [15-17]. In addition, the high-performance operation at low light intensity levels for indoor lighting condition is another advantage of organic solar cells in LSC device architecture [18-21], which have not been studied for IoT devices under different lighting conditions.

Here, we report a LSC architecture that is in-situ integrated with an array of bulk heterojunction organic photovoltaics. Among a wide variety of available fluorophores used as down-conversion materials such as organic dyes [22], lanthanide-based materials [23], and fluorescent proteins [24], we used quantum dots (QDs) due to high photoluminescence quantum yield (PLQY) and fine spectral photoluminescence tuning ability [25-32]. The photoluminescence of alloyed CdSe@ZnS QDs was synthetically engineered to match with the photovoltaic responsivity of P3HT:PC$_{61}$BM solar cells. Moreover, high PLQY of 91% and 64% in the synthesis batch and polymeric waveguide, respectively, facilitated an optical quantum efficiency of 44.7%. The fabricated organic solar cell-LSC architecture can generate sufficient power for the RFID tags to send signals in the high communication distance of 35 meters and ~0.1 km under the different lighting conditions of 0.1-sun and 1-sun conditions, respectively, which is sufficient for indoor and outdoor communications. LSCs integrated with organic solar cells possess potential to power-up independent IoT devices for data transmission, processing, and reception due to their scalable size, transparency, and configurable operation.

**EXPERIMENTAL DETAILS**

*Synthesis materials* - Cadmium oxide (>99.99% Aldrich), oleic acid (99% Alfa), zinc acetate dehydrate (>99.0% Sigma-Aldrich), selenium (>99.5% Aldrich), sulfur (Sigma-Aldrich), trioctylphosphine (90% Acros), 1-octanethiol (>98.5% Aldrich), ethanol and hexane solvents.

*Green-emitting QDs synthesis procedure* - We synthesized CdSe@ZnS QDs based on the previous method [33, 34] with some modifications. To achieve the green-emitting QDs with photoluminescence peak wavelength at 521 nm, 0.2 mmol CdO (0.026 g), 16 mmol OA (5 ml), and 4 mmol $Zn(ac)_2 \cdot 2H_2O$ (0.878 g) were mixed with each other in a 50 ml flask. The flask was heated to 120 °C in the schlenk line under constant purge of nitrogen gas. At 120 °C, the solution was purged by vacuum and nitrogen repeatedly. After the completion of evacuation process, the temperature was increased to 300 °C. After 1 hour, 0.2 mmol Se (0.016 g) and 3 mmol S (0.128 g) in 2 ml TOP was quickly introduced into the flask. After 10 minutes, 0.5 ml of 1-octanethiol was injected. After the injection, the heating mantle was removed to cool down the reaction. The obtained QDs were washed several times and re-dispersed in hexane solvent.

*Organic solar cell fabrication and characterization* - Inverted bulk heterojunction photovoltaic devices were fabricated on patterned indium tin oxide (ITO) glass substrates with the configurations of ITO/ZnO/P3HT:$PC_{61}BM$/$MoO_3$/Ag in the air. Firstly, ITO coated glass substrates were sonicated in 10% of Hellmanex liquid alkaline concentrate in de-ionized (DI) water for 15 minutes at 50 °C. The substrates were then rinsed with hot and cold DI water, respectively, and then sonicated in acetone for 15 minutes and isopropanol for 15 minutes, one after another. Next, the substrates were dried in the oven at 100 °C and treated with UV-ozone for 15 minutes. A sol-gel of ZnO (0.45 M) was prepared using zinc acetate dehydrate in 2-methoxyethanol and ethanolamine. The prepared solution was spin coated at 2000 rpm for 60 seconds to deposit ZnO thin film (~50 nm) on the substrate. A drying process was performed on a hot plate at 250 °C for 15 minutes at room temperature. The P3HT and $PC_{61}BM$ were used as it was received from Sigma-Aldrich. The P3HT:$PC_{61}BM$ blend was deposited by spin-coating a solution (20 mg.ml$^{-1}$) of P3HT:PC$_{61}$BM (1:0.6 ratio) in o-dichlorobenzene at 400 rpm for 150 seconds on top of ZnO layer. The P3HT:PC$_{61}$BM layer was dried at 150 °C on hot plate for 15 minutes to remove the excess solvent. The thickness of the active layer used in these devices is ∼ 210 nm. 10 nm of molybdenum trioxide (MoO$_3$), an electron blocking layer, was deposited on top of the active layer by thermal evaporation. A silver cathode (100 nm) was deposited by thermal evaporation through a shadow mask under a pressure of ∼ 2×10$^{-6}$ mbar to complete the device, resulting a device area of approximately 0.0256 cm$^2$. After the thermal evaporation, organic solar cells were immersed into PDMS and QD-PDMS mixture for curing.

*Current-voltage measurements* - Current-voltage (*J-V*) characteristics of organic solar cells were measured under an AM 1.5G illumination source with the intensity of 100 mW.cm$^{-2}$ in the ambient condition. The light intensity was adjusted with a NREL calibrated silicon solar cell. The Keithley 2400 source meter was used for the measurement of (*J-V*) characteristics of the organic photovoltaic devices. The current-voltage measurements were carried out for one individual organic solar cell pixel.

*Instrumentation and characterization* - We carried out photoluminescence and UV/Visible absorbance spectra of QDs by Edinburgh Instruments Spectroflourometer FS5 with Xenon lamp. The QDs solution was excited with 335 nm wavelength. The integrating sphere was used to measure absolute PLQY of the synthesized QDs. The optical fiber measurements was performed with Torus detector with 1.6 nm emission width of optical resolution in the range of 360-825 nm.

*Electro-optical measurements of the bulk heterojunction device* - In the electro-optical measurements of the bulk heterojunction device, one edge of the LSC was coupled with the organic solar cell and on the three other uncoupled edges, reflecting mirrors were placed in order to direct back the emitted light to the waveguide [35]. A mate black paper was placed at the back of the LSC in order to block any direct illumination from the solar simulator [36].

*RFID Chip Integration and Antenna Power Measurements* - For RFID transmitter and receiver chips, we used a 433 MHz WL101-341 RF receiver and DRA888TX ASK transmitter modules. We used 6 solar cell pixels to achieve the minimum operating voltage for these chips, ~1.5 V, even under light intensities as low as 0.1-sun. Moreover, to utilize voltage and current supply of each pixel, these 6 individual pixels were connected in series corresponding to a total voltage value of 3.51 V to obtain the required turn-on voltage for the RFID chips. An Arduino Uno R3 with an ATmega328P processor was utilized to code the transmitter and receiver modules to work at 8 Kbps data rate. The data transmission and receiving message were constantly monitored to ensure proper communication without data loss in the transmitted message. Moreover, the antenna power for transmitter module, which was powered by our solar cell-integrated LSC, was measured using a Rigol DSA815-TG spectrum analyzer that works within 9 kHz-1.5 GHz frequency range. During the experiments, the sun illumination intensity was changed from 0.1- to 1-sun and transmitted message and antenna power levels were recorded for each sun intensity.

**RESULTS AND DISCUSSIONS**

Different from the previous studies combining waveguides and solar cells in the post-LSC-fabrication phase [18-20], we directly integrated the photovoltaics inside the QD-doped polymeric waveguide during the waveguide formation. First, the synthesized QDs were mixed with the PDMS elastomer and curing agent and degassed under vacuum to remove bubbles. The fabricated organic solar cells were fixed at the edges of the pre-fabricated aluminum mold (Figure 1a). Then, the mixture was poured in the mold covering the organic solar cells (Figure 1b). The mold was placed in the oven and heated at the temperature of 70 °C for 6 hours to complete the curing process (Figure 1c). Upon completion of the curing process, the organic solar cells were attached to the QD-polymer composite. The QD-polymer composite integrated with the organic solar cells were removed from the mold (Figure 1d). The photograph of the device showed the sufficient level of transparency under ambient light (Figure 1e) and strong luminescence under UV radiation (Figure 1f). The direct integration of the waveguide and photovoltaic units advantageously enables the coupling of light from the low refractive index ($n_{PDMS}$=1.42) [37] to the high refractive index solar cells ($n_{opv}$=1.95) [38] without any reflection losses due to an air gap.

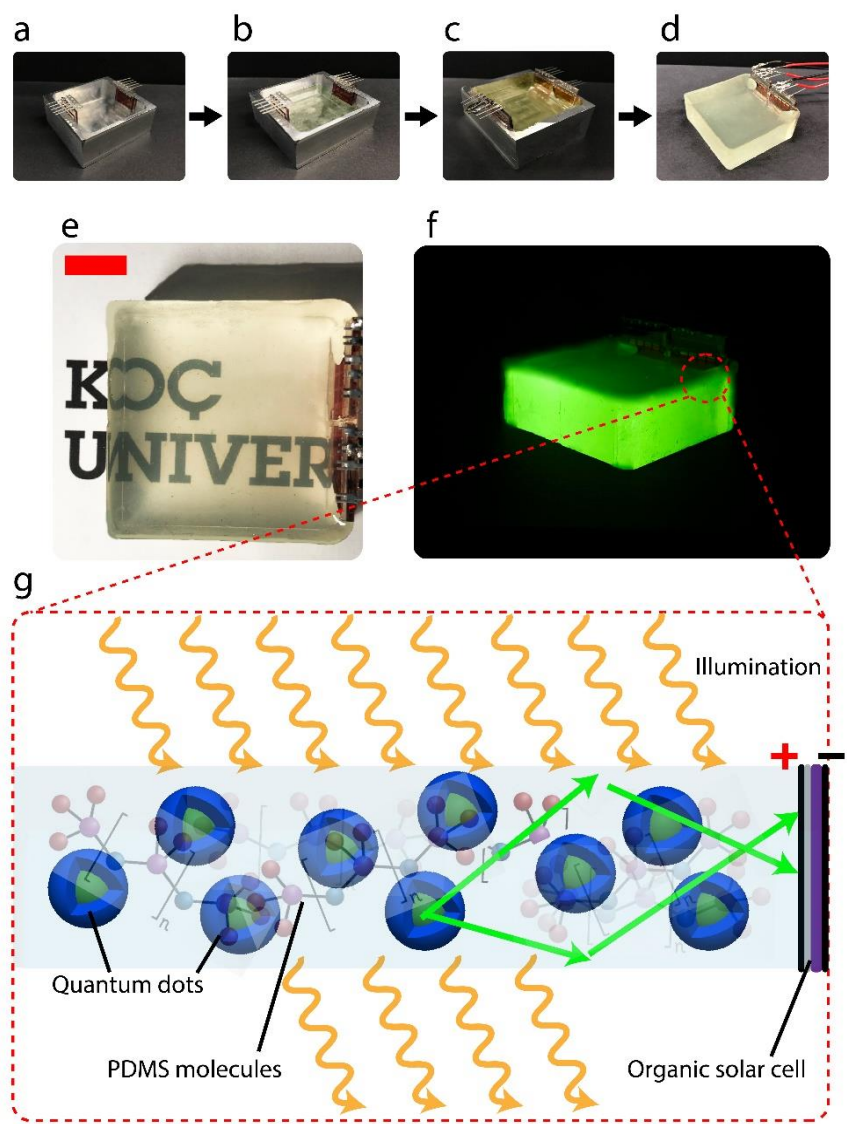

**Figure 1. (a)** The fabricated solar cells were placed to the edges of the 5×5×1 cm³ aluminum mold. **(b)** The green-emitting quantum dots and PDMS elastomer mixture was degassed and poured in the aluminum mold. **(c)** The mold was heated in the oven for 6 hours at 70 °C. **(d)** After the curing process was finished, the mold was taken out form the oven and the LSC was evacuated from the mold. **(e)** The photograph of the transparent device under ambient light (scale bar = 1 cm). **(f)** The quantum dot-based luminescent solar concentrator integrated with organic solar cells under UV illumination. **(g)** In the zoomed area shown from the edge, the generated photoluminescence from CdSe@ZnS quantum dots surrounded by PDMS molecules is absorbed by the integrated organic solar cell at the edge. Dark orange arrows represent the incoming and transmitted photons and green arrows represent the down-converted photons guided to the solar cells.

To maximize the energy harvesting efficiency, the photoluminescence of the fluorophore inside the waveguide needs to have good spectral overlap with the absorption of the solar cells. For that, the photoluminescence of the synthesized alloyed CdSe@ZnS QDs was tuned by varying the reaction time during the synthesis procedure and the photoluminescence was recorded at each reaction time. As the reaction time increased from 5 to 20 minutes, the peak wavelengths changed from 498 nm to 574 nm, respectively (Figure 2a). The green-emitting QDs with photoluminescence peak position at 521 nm, which has a size of 8.6 nm ± 0.97 nm (Figure 2b), was selected for the integration inside the polymeric matrix that matched with the absorbance peak wavelength of the active layer located at 515 nm (Figure 2a). At the same time, the green-emitting CdSe@ZnS QDs showed high PLQY of 91% and 64% in solution- and solid-state, respectively.

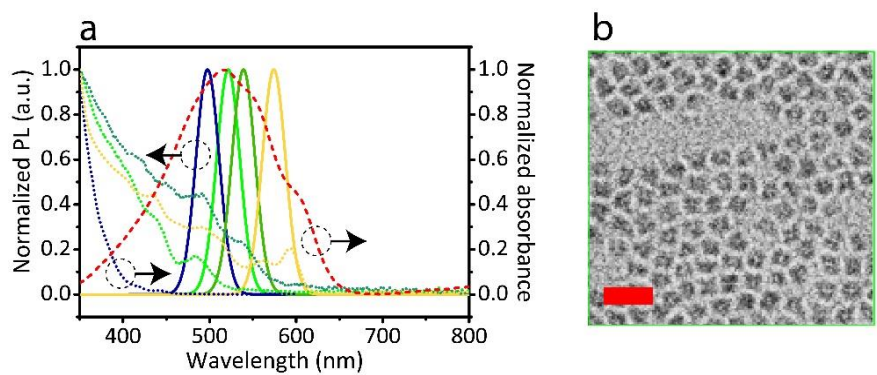

**Figure 2. (a)** The absorbance (dashed red line) of the organic active layer of the bulk heterojunction solar cell, the normalized absorbance (dashed lines), and photoluminescence (solid lines) of the quantum dots synthesized in different reaction times ranging from 5 to 20 minutes. As the reaction time elapses, the photoluminescence shows red shift. **(b)** The TEM image of the synthesized quantum dots (scale bar = 20 nm).

For light to electricity conversion, bulk heterojunction organic solar cells using P3HT as a donor and the fullerene derivative of $PC_{61}BM$ as an acceptor were fabricated in an inverted geometry of ITO/ZnO/P3HT:$PC_{61}BM$/$MoO_x$/Ag in the ambient condition. The P3HT:$PC_{61}BM$ blend was used as photoactive layer for LSC application because it is one of the most widely studied and successful organic semiconductor blend with proven record of efficiency and stability [39-41]. The energy band diagram was shown in Figure 3a, which confirms the suitability of the band alignment for charge transport and collection. The scanning electron microscopy (SEM) was used to confirm the thicknesses of various layers of the organic solar cell (Figure 3b). The current-voltage (*J-V*) characteristics was studied under 1-sun AM 1.5G illumination (Figure 3c). The solar cells achieved a power conversion efficiency (PCE) of 3.8 ± 0.36 % (mean ± SD, n = 6) with an open circuit voltage ($V_{oc}$) of 0.59 ± 0.07 V (mean ± SD, n = 6), a short-circuit current density ($J_{sc}$) of 11.9 ± 0.63 mA.cm$^{-2}$ (mean ± SD, n = 6) and a fill factor (FF) of 0.55. The achieved electrical parameters are well in the range of the state-of-the-art reports for the P3HT:$PC_{61}BM$-based organic solar cells [42].

To understand the effect of light intensity over the performance of the organic solar cells, the current-voltage (*J-V*) characteristics of the fabricated organic solar cells were investigated (Figure 3d). The systematic variation of $J_{sc}$ and $V_{oc}$ with respect to the illumination intensity were shown in Figures 2e and 2f after fitting with $R^2$ values of 0.987 and 0.995, respectively. Both the $J_{sc}$ and $V_{oc}$ values were increased by increasing the light intensity from 0.1- to 1-sun [43-45] and the slope of the $J_{sc}$ with respect to the light intensity in the log-log plot is linear, which reveals the single photon-induced processes [45, 46]. At the same time, the slope of the $V_{oc}$ with respect to light intensity in semi-log plot showed a linear variation for P3HT:PC$_{61}$BM solar cells (3e, f). Hence, the PCE of the organic solar cells are 2.8% for 0.1-sun and 3.8% under 1-sun illumination showing efficient operation within the intensity range.

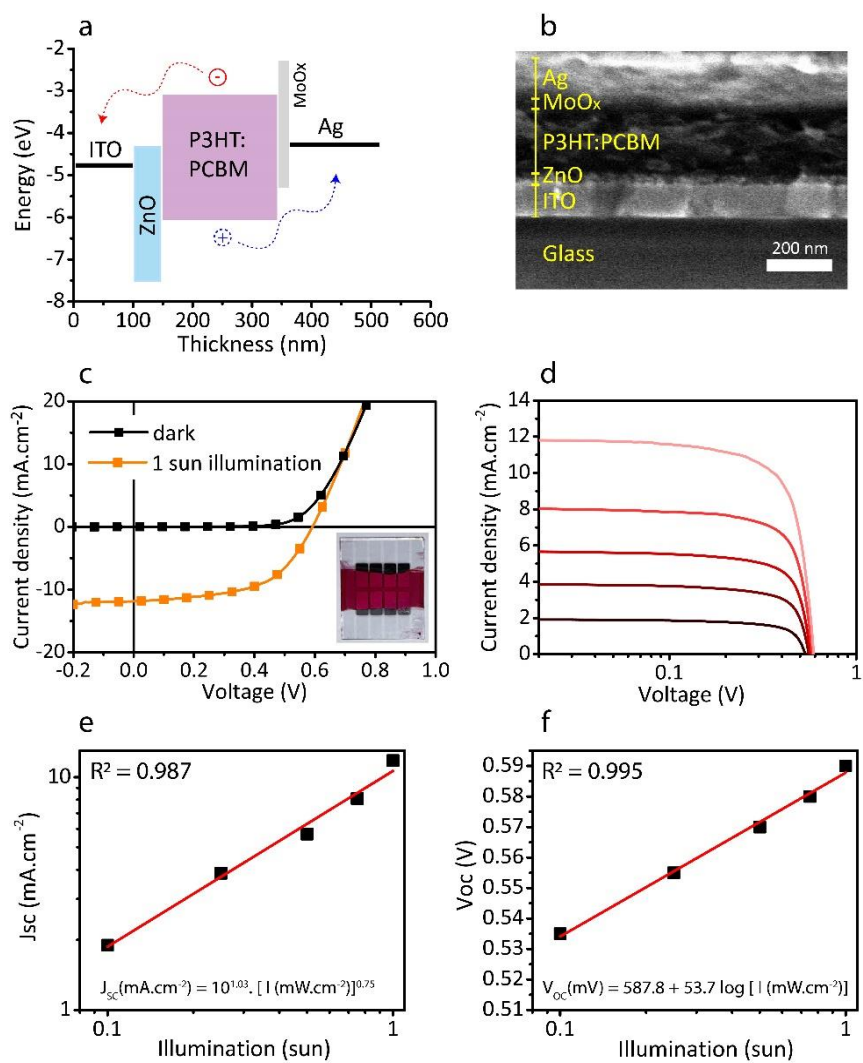

**Figure 3.** (**a**) Energy level diagram of inverted organic solar cell (ITO/ZnO/P3HT:PC$_{61}$BM/MoO$_x$/Ag) accompanied with charge transport mechanism. (**b**) Corresponding cross-sectional SEM image of the active area. (**c**) The (*J-V*) characteristics of the optimized single organic solar cell device under AM 1.5G illumination with 100 mW.cm$^{-2}$. Inset: Image of the fabricated device consisting of an array of 8 solar cell pixels. (**d**) (*J-V*), (**e**) Jsc, and (**f**) Voc characteristics of the organic solar cell at different light intensities ranging from 0.1- to 1-sun. The fitting line equations and the R$^2$ values were shown in each graph.

In order to investigate the optical properties of the fabricated waveguide integrated solar cell device, we measured the transmission of the QD doped slab and compared it with the bare PDMS waveguide with the similar thickness of 1 cm (Figure 4a). Upon the integration of QDs with high loading concentration of 0.8 wt%, the transmission level of the waveguide moderately modified due to the scattering and the absorption of QDs. The normalized photoluminescence and absorbance spectrum of the QDs inside the LSC architecture were also recorded as shown in Figure 4b. The photoluminescence peak position of the QDs showed a small change of 5 nm (from 521 nm to 526 nm) upon integration into the polymeric matrix. Moreover, the observed offset of 0.2 in the absorbance spectrum at 800 nm is mainly due to the light scattering from the surface of the PDMS layer and the scattering originated from the QDs at high loading concentration of 0.8 wt%, which was also shown in Figure 4a. Today, solid-state lighting sources that are used in internal lighting and in the mobile phones have strong blue spectral content [34, 47-49], and we investigated the performance of the organic solar cell and the integrated device under blue LED with the emission peak position at 445 nm and the power density of 100 mW.cm$^{-2}$ (Figure 4c). When the organic solar cell was being directly illuminated by the blue LED and the calibrated AM 1.5G light, it showed the $J_{SC}$ values of 14 mA.cm$^{-2}$ and 11.9 mA.cm$^{-2}$, respectively (Figure 4d).

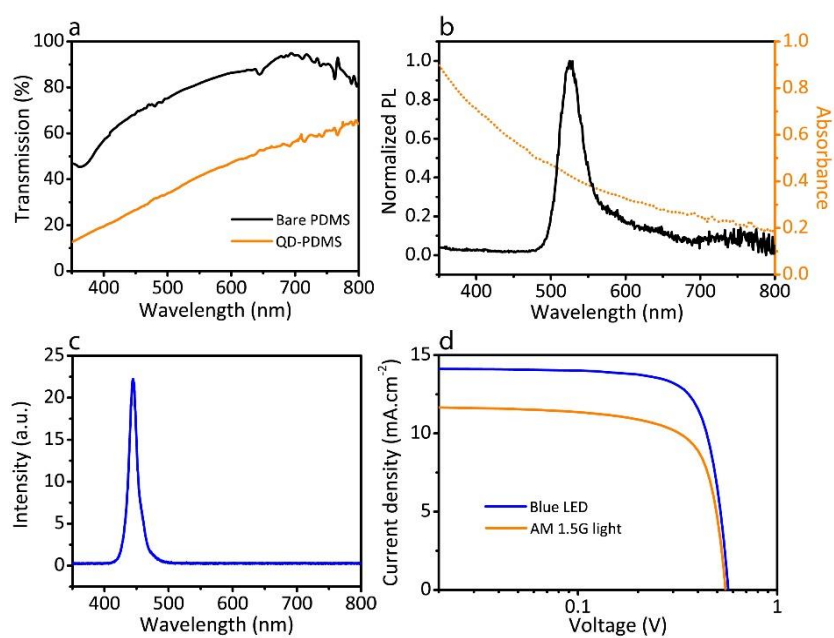

**Figure 4.** (a) The transmission spectra of the bare PDMS and QD-PDMS waveguides. (b) The normalized photoluminescence (black) and the absorbance (orange) spectrum of green-emitting QDs inside the LSC architecture. (c) The emission spectrum of the blue LED. (d) (*J-V*) curves of the fabricated organic solar cells under AM 1.5G light with the output power of 100 mW.cm$^{-2}$ and blue light with the output power of 100 mW.cm$^{-2}$.

In order to investigate the contribution of the illuminated light and the photoluminescence of the green-emitting QDs on the generated output power, we illuminated the waveguide surface with AM 1.5G light and recorded the output spectrum from the edge of the waveguide by using an optical fiber (Figure 5a). The spectral measurement from the edge of the integrated device revealed the QD photoluminescence peak positioned at 525 nm, which was compatible with the photoluminescence of the QDs inside the PDMS matrix (Figure 4b). At the same time, the spectral profile of the AM 1.5G light was also observed in the recorded edge emission spectrum, which was due to the total internal reflection of the scattered light inside the PDMS matrix. We also analyzed the optical performance of the integrated device under the standard AM 1.5G solar radiation. Due to the high PLQY of the synthesized QDs inside the polymeric matrix (64%), the optical quantum efficiency of the LSC was calculated as high as 44.7% based on Equation (1) as follows:

$$\eta_{OQE} = \frac{\int 1.05 \frac{\eta_{TIR}\eta_{PL}}{1+\beta\alpha(\lambda)L(1-\eta_{TIR}\eta_{PL})} \phi_{PL}(\lambda)d\lambda}{\int \phi_{PL}(\lambda)d\lambda} \quad (1)$$

In Equation (1), the optical quantum efficiency (*i.e.*, internal quantum efficiency) is defined as the fraction of the photons that reach the edges of the LSC to the absorbed photons. $\eta_{TIR}$ is the total internal reflection efficiency of the PDMS polymer calculated as 0.745 and $\eta_{PL}$ is the quantum efficiency of the synthesized QDs inside the polymeric matrix measured by using the integrating sphere as 0.64. L is the side length of the LSC equal to 5 cm and β was considered as 1.4 based on reference [50]. $\phi_{PL}(\lambda)$ and $\alpha(\lambda)$ are the photoluminescence spectrum and the absorption coefficient of QDs over the entire wavelength region [50-53]. The high value of optical quantum efficiency of the fabricated devices (44.7%) results in the concentration factor of 2.2 based on C.F = $\eta_{OQE}$ × G [54], which indicates the effective light concentration inside the QD-LSC architecture. The optical efficiency of the integrated device (*i.e.*, the external optical efficiency) is measured based on Equation (2), which defines the ratio of the photons absorbed by the integrated OSC device to the photons impinging on the waveguide surface.

$$\eta_{opt} = \frac{I_{LSC}}{I_{OPV} \times G} \quad (2)$$

In Equation (2), $I_{LSC}$ is the short circuit current level of organic solar cell when the top surface of the LSC is being illuminated by the light, $I_{OPV}$ is the short circuit current level of organic solar cell when being directly illuminated by the light (Figure 4d) and G is the optical gain factor of the LSC calculated based on G = $A_{top\ surface}$/$A_{collection\ edge}$ equal to 5. When being illuminated by a calibrated AM 1.5G solar spectrum, the short circuit current of the integrated device was measured as 2.9 mA.cm$^{-2}$ (Figure 5b), which yielded to the optical efficiency of 4.9%.

The optical investigation of the bulk heterojunction LSC device integrated with organic solar cell showed that when the integrated device was being illuminated by the blue LED, a current density of 3.5 mA.cm$^{-2}$ was observed (Figure 5c), which corresponded to the optical efficiency of 5%. To investigate the effect of emission from QDs on the enhancement of the output short-circuit current in the integrated device, another device integrated with organic solar cell was fabricated by using only PDMS polymer without integration of the green-emitting QDs. Under the same light intensity level, a 2.2-fold lower short-circuit current level of 1.6 mA.cm$^{-2}$ was obtained (Figure 5c). Here, the significant level of current by the pure PDMS is induced due to the large divergence angle of LED that are guided to the solar cell. In principle, the LSC thickness can decrease down to hundreds of microns with constant illumination area and the loading concentration. However, this will result in the increase of the G factor, which consequently leads to the significant reduction in the optical efficiency of the device. Employing the LSC architecture may result in the lower output power from the bulk heterojunction in comparison with the direct illumination of the solar cells. However, the high transparency in the context of transparent electronics and capability of light collection over large area as solar windows in indoor and outdoor applications will outcompete the generated power from the direct illumination in the prolonged time.

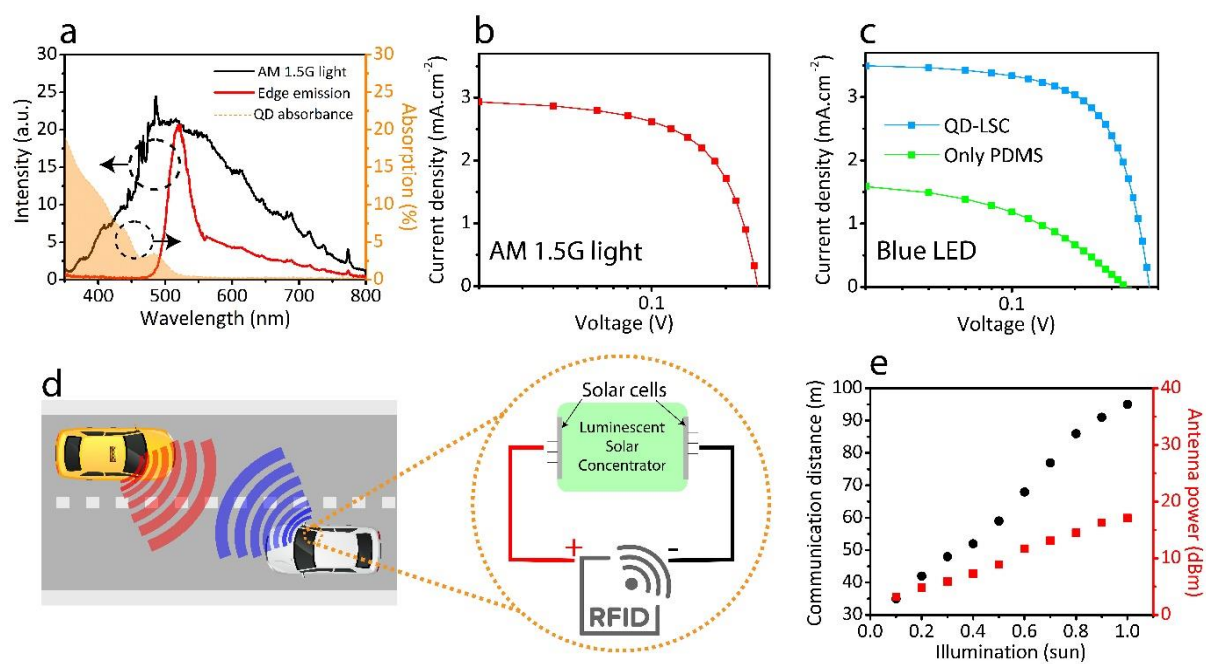

**Figure 5.** (**a**) The absorption percentage of the QDs (orange), the solar spectrum (black), and the edge emission (red) of QD-LSC under AM 1.5G light at visible wavelength region. (**b**) The (*J-V*) curves of the solar cell-integrated LSC under AM 1.5G light with the output power of 100 mW.cm$^{-2}$. (**c**) (*J-V*) curves of the solar cell-integrated LSC under blue light with the output power of 100 mW.cm$^{-2}$. (**d**) The schematic represents the organic solar cell-LSC architecture, which empowers a RFID chips to be used for connected cars applications. (**e**) The effect of illumination power on the maximum communication distance (black dots) and the antenna power (red dots) of the RFID chip.

Five main IoT technologies including radio frequency identification (RFID), wireless sensor networks (WSN), middleware, cloud computing, and IoT application software are widely applied for the implementation of IoT products and services [55]. Among them, we selected RFID, which is an automatic technology that aids machines or computers to identify objects, record metadata and control individual target through radio waves [56, 57]. A typical RFID system consists of a transmitter/receiver microchip connected to an antenna, which can be attached to an object as the identifier. Active RFID tags have already started to be used for vehicle positioning (Figure 5d), automobile connections, manufacturing, hospital laboratories, and remote-sensing (e.g., temperature, pressure, and chemical) [55, 58]. Active RFID tags have their own battery supply and we replaced this unit with our fabricated LSC (Figure 5d). We tested transmitter performance by changing illumination intensity from 0.1- to 1-sun condition to evaluate the RFID functionality. The experiment was carried out without any additional antenna. For RFID part, a commercial 433 MHz DRA888TX ASK RF transmitter module with optimal operating voltage of 1.5 V and injection current of 2.1 mA was used.

The maximum communication distance, which corresponds to the maximum distance to communicate with 8 Kbps rate without any information loss, was measured for different sun intensities. The transmitter power was measured with a spectrum analyzer. The results indicated that output antenna power varies between 3-17 dBm, which corresponds to maximum communication distances between approximately 35 to 95 meters for 0.1- and 1-sun illumination, respectively (Figure 5e). The communication distance results revealed that fabricated device can be used for RFID applications for connected cars even with a RF transceiver chip that has a built-in low power antenna. Although up to 95 meters communication distance can be used for short range vehicle-to-vehicle communication, the communication distance can be improved by utilizing additional transmitter and receiver antennas, which only requires additional solar cell pixels to be powered. Moreover, the high current levels under blue light (Figure 5c) is also sufficient for supplying enough energy by solid-state lighting sources. Our demonstration can be directly applied for vehicle-to-vehicle communications that prevents car crashes in short range traffic or smart parking lot purposes, and it showed the potential of LSCs integrated with organic solar cells for future IoT applications. A broad range of low power communication modules can be directly integrated to our system [59]. Moreover, utilizing low power GSM modules with 5G technology may bring the opportunity to create vehicle-to-cloud or even vehicle-to-everything communication and our design idea can be utilized efficiently in large area vehicles like trains or ferries as well. Due to the transparency and flexibility of LSC polymer, LSCs can be further integrated on the windows and other surfaces of the vehicles [60].

**CONCLUSIONS**

In this study, a single integrated unit of bulk heterojunction solar cell array and quantum dot luminescent solar concentrator was designed, fabricated, and used for data transmission for IoT. The optical properties of the synthesized CdSe@ZnS QDs were tailored to match with the absorbance peak of the P3HT: $PC_{61}BM$ as the active layer of the organic photovoltaics. The synthesized QDs showed high PLQY of 64% inside the polymer that led to an optical quantum efficiency of 44.7%. The tailoring of optical properties and the direct integration of the organic solar cell inside the QD-LSC architecture further led to the optical efficiency of 4.9% and 5.0% under AM 1.5G and blue LED illumination, respectively. The fabricated QD-LSC integrated with organic photovoltaics was used to empower the RFID chips for connecting cars in IoT applications. The generated power by 0.1-sun illumination led to the communication distance of up to 35 meters. By increasing the illumination intensity up to 1-sun illumination, the communication distance can reach up to ~0.1 km, which is suitable for all types of vehicle-to-vehicle communication, traffic and parking lots to avoid crashes and optimized navigation. Therefore, this study reporting a new transparent-electronic approach for empowering IoT holds promise for wireless communication of independent devices.

## ACKNOWLEDGEMENTS

S.N. acknowledges the Turkish Academy of Sciences (TÜBA-GEBİP) and Science Academy (BAGEP). The authors thank the Koç University Tüpraş Energy Center (KUTEM).

**Conflicts of Interests**

Authors declare no competing financial interests.

[46] Photocurrent and Photovoltage Study," *Advanced Functional Materials,* vol. 21, pp. 1419-1431, 2011/04/22 2011.

[47] S. Sadeghi, B. G. Kumar, R. Melikov, M. M. Aria, H. B. Jalali, and S. Nizamoglu, "Quantum dot white LEDs with high luminous efficiency," *Optica,* vol. 5, pp. 793-802, 2018.

[48] E. Jang, S. Jun, H. Jang, J. Lim, B. Kim, and Y. Kim, "White-light-emitting diodes with quantum dot color converters for display backlights," *Advanced materials,* vol. 22, pp. 3076-3080, 2010.

[49] S. Sadeghi, S. E. Mutcu, S. B. Srivastava, G. Aydindogan, S. Caynak, K. Karslı, *et al.*, "High quality quantum dots polymeric films as color converters for smart phone display technology," *Materials Research Express,* vol. 6, p. 035015, 2018.

[50] V. I. Klimov, T. A. Baker, J. Lim, K. A. Velizhanin, and H. McDaniel, "Quality factor of luminescent solar concentrators and practical concentration limits attainable with semiconductor quantum dots," *ACS Photonics,* vol. 3, pp. 1138-1148, 2016.

[51] K. Wu, H. Li, and V. I. Klimov, "Tandem luminescent solar concentrators based on engineered quantum dots," *Nature Photonics,* vol. 12, p. 105, 2018.

[52] H. Zhao, Y. Zhou, D. Benetti, D. Ma, and F. Rosei, "Perovskite quantum dots integrated in large-area luminescent solar concentrators," *Nano energy,* vol. 37, pp. 214-223, 2017.

[53] H. Zhao, D. Benetti, L. Jin, Y. Zhou, F. Rosei, and A. Vomiero, "Absorption Enhancement in "Giant" Core/Alloyed-Shell Quantum Dots for Luminescent Solar Concentrator," *Small,* vol. 12, pp. 5354-5365, 2016.

[54] M. Wei, F. P. G. de Arquer, G. Walters, Z. Yang, L. N. Quan, Y. Kim*, et al.*, "Ultrafast narrowband exciton routing within layered perovskite nanoplatelets enables low-loss luminescent solar concentrators," *Nature Energy,* p. 1, 2019.

[55] I. Lee and K. Lee, "The Internet of Things (IoT): Applications, investments, and challenges for enterprises," *Business Horizons,* vol. 58, pp. 431-440, 2015.

[56] X. Jia, Q. Feng, and C. Ma, "An efficient anti-collision protocol for RFID tag identification," *IEEE Communications Letters,* vol. 14, pp. 1014-1016, 2010.

[57] X. Jia, Q. Feng, T. Fan, and Q. Lei, "RFID technology and its applications in Internet of Things (IoT)," in *2012 2nd international conference on consumer electronics, communications and networks (CECNet)*, 2012, pp. 1282-1285.

[58] J. Wang, D. Ni, and K. Li, "RFID-based vehicle positioning and its applications in connected vehicles," *Sensors,* vol. 14, pp. 4225-4238, 2014.

[59] N. Akhtar, S. C. Ergen, and O. Ozkasap, "Vehicle mobility and communication channel models for realistic and efficient highway VANET simulation," *IEEE Transactions on Vehicular Technology,* vol. 64, pp. 248-262, 2014.

[60] C. Ding, H. Li, W. Zheng, Y. Wang, N. Chang, and X. Lin, "Luminescent solar concentrator-based photovoltaic reconfiguration for hybrid and plug-in electric vehicles," in *2016 IEEE 34th International Conference on Computer Design (ICCD)*, 2016, pp. 281-288.
16